\documentclass[paper]{elsarticle}
\usepackage{epstopdf}
\usepackage{epsfig}
\usepackage{color,framed}
\usepackage{amsmath,amsthm,amssymb}
\usepackage{hyperref}
\usepackage{lineno}
\usepackage{cases}
\usepackage{graphicx}
\usepackage{dcolumn,color}

\begin{document}

\begin{frontmatter}

\title{Linearization of a warped $f(R)$ theory in the higher-order frame II: the equation of motion approach}

\author[YZaddress]{Yuan Zhong }
\address[YZaddress]{School of Science, Xi'an Jiaotong University, Xi'an 710049, People's Republic of China}
\fntext[myfootnote]{Corresponding author.}

\author[KYaddress]{Ke Yang }
\address[KYaddress]{School of Physical Science and Technology,
Southwest University, Chongqing 400715, People's Republic of China}
\author[Liuaddress]{Yu-Xiao Liu\fnref{myfootnote} }
\ead[url]{liuyx@lzu.edu.cn}
\address[Liuaddress]{Institute of Theoretical Physics, Lanzhou University,\\
 Lanzhou 730000, People's Republic of China}

\begin{abstract}
Without using conformal transformation, a simple type of five-dimensional $f(R)-$brane model is linearized directly in its higher-order frame. In this paper, the linearization is conducted in the equation of motion approach.  We first derive all the linear perturbation equations without specifying a gauge condition. Then by taking the curvature gauge we derive the master equations of the linear perturbations. We show that these equations are equivalent to those obtained in the quadratical action approach [Phys. Rev. D 95 (2017) 104060], except the vector sector, in which a constraint equation can be obtained in the equation of motion approach but absent in the quadratical action approach. Our work sets an example on how to linearize higher-order theories without using conformal transformation, and might be useful for studying more complicated theories.
\end{abstract}

\begin{keyword}
$f(R)$ gravity \sep linear perturbations\sep warped extra dimensions
\end{keyword}

\end{frontmatter}


\section{Introduction}
In the last two decades, warped extra dimensions have been applied to explain the large hierarchy between the electroweak scale and the Planck scale~\cite{RandallSundrum1999,CabrerGersdorffQuiros2010,CabrerGersdorffQuiros2011,CabrerGersdorffQuiros2011,RaychaudhuriSridhar2016}, the splitting of fermion masses~\cite{GherghettaPomarol2000}, the reproduction of Newtonian gravity on a lower-dimensional hypersurface~\cite{RandallSundrum1999a,Gremm2000,DeWolfeFreedmanGubserKarch2000,CsakiErlichHollowoodShirman2000}, and recently the LHC diphoton excess~\cite{MegiasPujolasQuiros2016} and LHCb anomalies~\cite{MegiasPanicoPujolasQuiros2016} (see~\cite{Quiros2015,Ponton2012,Liu2017} for recent reviews on the theory and phenomenology of warped spaces).

In a type of warped extra dimensional model, our world is described as a four-dimensional topological domain wall generated by a background scalar field in Einstein's gravity~\cite{RubakovShaposhnikov1983,Gremm2000,DeWolfeFreedmanGubserKarch2000,CsakiErlichHollowoodShirman2000}. But it is also possible to generate pure geometric domain wall solutions in $f(R)$ theory~\cite{ZhongLiu2016}, where the gravitational Lagrangian is an arbitrary function of the scalar curvature (see~\cite{Starobinsky1980,BarrowOttewill1983,NojiriOdintsov2003d,CarrollDuvvuriTroddenTurner2004,CapozzielloCarloniTroisi2003,NojiriOdintsov2006} for early literatures and~\cite{SotiriouFaraoni2010,DeTsujikawa2010} for comprehensive reviews on $f(R)$ theory and its cosmological phenomenology). In this case, the domain wall is non-topological, because it connects two equivalent anti-de Sitter vacuum. More $f(R)$ domain wall solutions can be found in Refs. ~\cite{ZhongLiu2016,ParryPichlerDeeg2005,AfonsoBazeiaMenezesPetrov2007,HoffdaSilvaDias2011,LiuZhongZhaoLi2011,BazeiaMenezesPetrovSilva2013,BazeiaLobaoMenezesPetrovSilva2014,XuZhongYuLiu2015,YuZhongGuLiu2015}.

It is both important and interesting to study the linearization of domain wall solutions in a warped $f(R)$ gravity. Because the linearization not only tells us whether a solution is stable against small metric perturbation, but also offers the spectra of graviton and radion, which is important for phenomenological applications. As a higher-order curvature theory, $f(R)$ gravity might have some new features than Einstein's theory. But a direct linearization of $f(R)$ domain wall is not easy, not only because one needs to carefully eliminate the residual gauge degrees of freedom, but also because the equation of motion in $f(R)$ gravity contains derivative up to fourth order. In literature, one usually rewrite the fourth-order $f(R)$ (the higher-order frame) as a second-order Einstein-scalar theory (the Einstein frame) by introducing a proper conformal transformation~\cite{BarrowCotsakis1988,Maeda1989,Wands1994,CapozzielloRitisMarino1997,FaraoniGunzigNardone1999}. Therefore, to linearize a $f(R)$ domain wall, one can first do the conformal transformation and then conduct the linearization in the Einstein frame~\cite{ZhongLiu2016}. To the authors' knowledge, there is only a few works directly confront the linearization of $f(R)$ theory without using conformal transformation (see~\cite{HwangNoh1996} for an example in $f(R)$ cosmology). If two frames are equivalent, the perturbation equations must be frame independent. But this conclusion is not obvious. Most importantly, when more general higher-order curvature theories are considered, the conformal transformation might not be convenient any more, then a direct analysis in the higher-order frame is inevitable. The aim of this work is to confront the linearization of $f(R)$ gravity with a warped geometry in the higher-order frame. In a previous work~\cite{ZhongLiu2017}, the linearization of $f(R)$ domain wall has been conducted in the quadratical action approach. In this paper, we redo the task in the equation of motion approach, and compare the results with those in Ref.~\cite{ZhongLiu2017}.

This paper is organized as follows. In next section, we briefly review the model and specify some conventions. The linearization of warped $f(R)$ domain walls is conducted in Sec.~\ref{sec3}, where the metric perturbation is decomposed into scalar, tensor and vector parts. The gauge degrees of freedom will not be fixed until in Sec.~\ref{sec4}, where the curvature gauge will be applied to simplify the scalar perturbation equation.  The result is summarized in Sec.~\ref{secSum}.
\section{The model}
In this paper, we consider a five-dimensional metric $f(R)$ gravity
\begin{eqnarray}
\label{action'}
S=\frac{1}{2\kappa_5^2}\int d^5x\sqrt {-g}f(R).
\end{eqnarray}
The corresponding Einstein field equations are
\begin{eqnarray}
\label{eqEE}
R_{MN}f_R-\frac12g_{MN}f(R)+(g_{MN}\hat{\square}^{(5)}
-\nabla_M\nabla_N)f_R=0,
\end{eqnarray} where $\hat{\square}^{(5)}=g^{MN}\nabla_{M}\nabla_{N}$ denotes the five-dimensional d'Alembertian operator defined by the metric $g_{MN}$ and the covariant derivative $\nabla_M$. The capital letters $M,N=0,1,2,3,5$ represent the bulk indices, and $f_R\equiv df(R)/dR$.

A warped space is described by the following metric:
\begin{eqnarray}
\label{metric}
ds^2=a^2(r)\eta_{MN}dx^M dx^N,
\end{eqnarray}
where $\eta_{MN}=\textrm{diag}(-1,1,1,1,1)$ and $a(r)$ is the warp factor, which depends only on the extra dimension $r\equiv x^5$.
Given the line element \eqref{metric}, it is easy to write the expressions of the connection, the Ricci tensor, the Ricci scalar and the last two terms in Eq.~\eqref{eqEE}:
\begin{eqnarray}
\label{backquantities1}
\Gamma^P_{MN}&=& 2\frac{{\delta _{(M}^P{\partial _{N)}}a}}{a} - {\eta _{MN}}\frac{{{\partial ^P}a}}{a},\\
R_{MN}&=&6\frac{\partial_{M}a\partial_{N}a}{a^2}
-3\frac{\partial_{M}\partial_{N}a}{a}
-2\eta_{MN}\left(\frac{a'}{a}\right)^2
-\eta_{MN}\frac{a''}{a},\\
R&=&-4 a^{-2}\left[\left(\frac{a'}{a}\right)^2+2\frac{a''}{a}\right],\\
\nabla_M\nabla_Nf_R
&=& {\partial _M}{\partial _N}f_R - 2\frac{{\delta _{(M}^r{\partial _{N)}}a}}{a}f_R'
+ {\eta _{MN}}f_R'\frac{{a'}}{a},\\
\label{backquantities5}
g_{MN}\hat{\square}^{(5)} f_R
&=&{\eta _{MN}}({\square ^{(5)}}f_R + 3f_R'\frac{{a'}}{a})={\eta _{MN}}( f_R'' + 3f_R'\frac{{a'}}{a}).
\end{eqnarray}
Here, we have used the following notations:
\begin{enumerate}
	\item The primes represent derivatives with respect to
	the extra dimension $r$.
	\item The bulk indices are always raised and lowered in terms of $\eta^{MN}$ and $\eta_{MN}$, respectively. For example, in the last term of $\Gamma^P_{MN}$ we used $\partial^{P}\equiv\eta^{PQ}\partial_{Q}$.
	\item $\square^{(5)}\equiv\eta^{MN}\partial_M\partial_N\equiv\partial_M\partial^M$, and ${\square^{(4)}}\equiv\eta^{\mu\nu}\partial_\mu
	\partial_\nu\equiv\partial_\mu\partial^\mu$ are d'Alamber operators defined with Minkowski metrics and the ordinary partial derivatives. Obviously, $\square^{(5)}=\square^{(4)}+\partial_r\partial_r$.
	\item The symmetrization bracket of tensor indices is defied as
	\begin{eqnarray}
	T^P_{~Q(M_1M_2\cdots M_n)}\equiv\frac{1}{n!}\left(T^P_{~QM_1M_2\cdots M_n}+\textrm{permutations of}~M_1,M_2,\cdots, M_n\right),
	\end{eqnarray}
	for example,
	\begin{eqnarray}
	\frac{{\delta _{(M}^r{\partial _{N)}}a}}{a}
	\equiv\frac12\left(\frac{{\delta _{M}^r{\partial _{N}}a}}{a}
	+\frac{{\delta _{N}^r{\partial _{M}}a}}{a}\right).
	\end{eqnarray}
\end{enumerate}
Substituting Eqs.~\eqref{backquantities1}-\eqref{backquantities5} into the Einstein field equations \eqref{eqEE}, we get
\begin{eqnarray}
\label{backEinstein}
&&{\eta _{MN}}\left( - \frac{1}{2}{a^2}f(R) + f_R'' - 2f_R{\left( {\frac{{a'}}{a}} \right)^2} + 2f_R'\frac{{a'}}{a} - f_R\frac{{a''}}{a} \right)\nonumber\\
&&+ 6f_R\frac{{{\partial _M}a{\partial _N}a}}{{{a^2}}} - 3f_R\frac{{{\partial _M}{\partial _N}a}}{a} - {\partial _M}{\partial _N}f_R + 2\frac{{\delta _{(M}^r{\partial _{N)}}a}}{a}f_R' = 0.
\end{eqnarray}
Obviously, the non-trivial components are
\begin{eqnarray}
\label{Einstein1}
a^2
f(R)+\left[4f_R\left(\frac{a'}{a}\right)^2+2f_R\frac{a''}{a}\right]
-4f'_R\frac{a'}{a}-2f''_R=0,
\end{eqnarray}
and
\begin{eqnarray}
8f_R\left(\frac{a'}{a}\right)^2-8f_R\frac{a''}{a}+8f'_R\frac{a'}{a}
-a^2f(R)=0.
\end{eqnarray}
The summation of the above two equations leads to another useful identity:
\begin{eqnarray}
\label{EinsteinEq3}
{ - 2 f_R^\prime \frac{{a'}}{a} + 3{f_R}\frac{{a''}}{a} - 6{f_R}{{\left( {\frac{{a'}}{a}} \right)}^2} + f_R^{\prime \prime }}  = 0,
\end{eqnarray}
which is a second-order differential equation for $f_R(r)$, and can be analytically solved if the warped factor is simple enough, see Ref.~\cite{{ZhongLiu2016}} for some examples.  
\section{The linearization of Einstein equations}
\label{sec3}
Once the background solution is obtained, the next step is to consider the linear stability of the solution against small linear perturbation. Let us assume that the background solution is $\{a(r), f_R(r)\}$, and the metric perturbation is 
\begin{eqnarray}
\delta g_{MN}=a(r)^2 h_{MN}(x^{\mu},~r),
\end{eqnarray}
then the total metric reads
\begin{eqnarray}
g_{MN}=a(r)^2[\eta_{MN}+h_{MN}].
\end{eqnarray}
By using the orthogonal relation of the total metric
\begin{eqnarray}
g^{MP}g_{PN}=\delta^M_{~N},
\end{eqnarray}
one immediately concludes that, to the linear order
\begin{eqnarray}
\delta g^{MN}=-a(r)^{-2} h^{MN}, \quad h^{MN}\equiv \eta^{MP}\eta^{NQ}h_{PQ}.
\end{eqnarray}
Similarly, one can derive the linear perturbations of the connection, the Ricci tensor and the scalar curvature:
\begin{eqnarray}
\label{pertGamma}
\delta \Gamma^P_{MN}&=&\partial_{(M} h^P_{N)}
-h_{MN}\frac{\partial^P a}{a}
-\frac12\partial^P h_{MN}
+{\eta}_{MN}\frac{a'}{a}h^{P}_r,\\
\label{pertRicci}\delta R_{MN}&=&\partial_P\partial_{(M}h^P_{N)}
-\frac12\square^{(5)} h_{MN}
-\frac32\frac{a'}{a}\partial_r h_{MN}
-\frac{a''}{a}h_{MN}
-2\left(\frac{a'}{a}\right)^2 h_{MN}
\nonumber\\
&+&\eta_{MN}\frac{a'}{a}\partial_P h^{P}_r
+\eta_{MN}\frac{a''}{a}h_{rr}
+2\eta_{MN}\left(\frac{a'}{a}\right)^2h_{rr}
-\frac12\partial_{M}\partial_{N}h
\nonumber\\
&-&\frac12\eta_{MN}\frac{a'}{a}\partial_r h
+3\frac{{a'}}{a}{\partial _{(M}}{h_{N)r}} - 2\frac{{a'}}{a}\frac{{{\partial _{(M}}a{h_{N)r}}}}{a}
+2\left(\frac{a'}{a}\right)^2\eta_{r(M}h_{N)r},\\
\label{pertR}
a^{2}\delta R&=&\partial_M\partial_N h^{MN}
-\square^{(5)} h
-4\frac{a'}{a} h'
+8\frac{a'}{a}\partial_P h^P_r
+8\frac{a''}{a}h_{rr}
+4\left(\frac{a'}{a}\right)^2h_{rr}.\quad\quad
\end{eqnarray}
Here $h$ is defined as $h\equiv\eta^{MN}h_{MN}$. 

The task of linearization is to derive the master equations for $h_{MN}$.
In the equation of motion approach, the linear perturbation equations are obtained by perturbing all terms in the Einstein equations:
\begin{eqnarray}
&&\delta R_{MN}f_R+R_{MN}\delta f_R-\frac12\delta g_{MN}f(R)
-\frac12g_{MN}f_R\delta R\nonumber\\
&+&\delta(g_{MN}\hat{\square}^{(5)} f_R)-\delta(\nabla_M\nabla_Nf_R)=0.
\label{eqPertubedEE}
\end{eqnarray}
The last two terms can be expanded as
\begin{eqnarray}
\delta(\nabla_M\nabla_N f_R)&=&(\partial_M\partial_N-\Gamma^P_{MN}\partial_P)\delta f_{R}-\delta \Gamma^P_{MN}\partial_P f_R,\\
\delta(g_{MN}\hat{\square}^{(5)} f_R)&=&\delta g_{MN}\hat{\square}^{(5)} f_R
+g_{MN} \delta g^{PQ}(\nabla_P\nabla_Q f_R)\nonumber\\
&+&g_{MN}g^{PQ}\delta(\nabla_P\nabla_Q f_R),
\end{eqnarray}
or more explicitly,
\begin{eqnarray}
\label{lastTerm1}
\delta ({\nabla _M}{\nabla _N}{f_R})
&= &{\partial _M}{\partial _N}\delta {f_R}
- 2\frac{{{\partial _{(M}}a{\partial _{N)}}\delta {f_R}}}{a}
+ {\eta _{MN}}\frac{{a'}}{a}{\partial _r}\delta {f_R}+ {f_R^{\prime}}\frac{{a'}}{a}{h_{MN}}  \nonumber\\
& -& {f_R^{\prime}}{\partial _{(M}}{h_{N)r}}
 + \frac{1}{2}{f_R^{\prime}}{\partial _r}{h_{MN}}
- {\eta _{MN}}{f_R^{\prime}}\frac{{a'}}{a}{h_{rr}},\\
\label{lastTerm2}
\delta(g_{MN}\hat{\square}^{(5)} f_R)
&=&
{h_{MN}}(f_R'' + 3\frac{{a'}}{a}f_R')
+{\eta _{MN}}\big(\square^{(5)} \delta f_R + 3\frac{{a'}}{a}{\partial _r}\delta f_R \nonumber\\
&- & f_R'{\partial ^P}{h_{Pr}} - 3f_R'\frac{{a'}}{a}{h_{rr}} + \frac{1}{2}f_R'{\partial _r}h - f_R''{h_{rr}}\big).
\end{eqnarray}

Plugging Eqs.~\eqref{pertRicci}-\eqref{pertR} and \eqref{lastTerm1}-\eqref{lastTerm2} into Eq.~\eqref{eqPertubedEE},  after a long but straightforward calculation, we finally obtain the tensor form of the linearized Einstein equations:
\begin{eqnarray}
\label{generalEQ}
&&- \frac{1}{2}f_R\square^{(5)} {h_{MN}}
- \frac{3}{2}f_R\frac{{a'}}{a}{\partial _r}{h_{MN}}
- \frac{1}{2}f_R'{\partial _r}{h_{MN}}
- \frac{1}{2}f_R{\partial _M}{\partial _N}h
- {\partial _M}{\partial _N}\delta f_R \nonumber\\
&&
+ f_R{\partial _P}{\partial _{(M}}h_{N)}^P
+ 3f_R\frac{{a'}}{a}{\partial _{(M}}{h_{N)r}}
- 2f_R\frac{{a'}}{a}\frac{{{\partial _{(M}}a{h_{N)r}}}}{a}
+ 2f_R{\left( {\frac{{a'}}{a}} \right)^2}{\eta _{r(M}}{h_{N)r}}\nonumber\\
&&
+ 6\frac{{{\partial _M}a{\partial _N}a}}{{{a^2}}}\delta f_R
- 3\frac{{{\partial _M}{\partial _N}a}}{a}\delta f_R
+ 2\frac{{{\partial _{(M}}a{\partial _{N)}}\delta f_R}}{a}
+ f_R'{\partial _{(M}}{h_{N)r}} \nonumber\\
&&
+ {\eta _{MN}}\mathcal{I}
= 0,
\end{eqnarray}
where the scalar $\mathcal{I}$ is defined as
\begin{eqnarray}
\mathcal I&=&
\square^{(5)} \delta f_R + \frac{1}{2}f_R\square^{(5)} h
- \frac{1}{2}f_R{\partial _M}{\partial _N}{h^{MN}}
- 3f_R\frac{{a'}}{a}{\partial _M}h_r^M
- 3f_R\frac{{a''}}{a}{h_{rr}} \nonumber\\
&+&
\frac{3}{2}f_R\frac{{a'}}{a}{\partial _r}h
- 2f_R'\frac{{a'}}{a}{h_{rr}}
- f_R'{\partial ^M}{h_{Mr}}
+ \frac{1}{2}f_R'{\partial _r}h
- f_R''{h_{rr}}
+ 2\frac{{a'}}{a}{\partial _r}\delta f_R \nonumber\\
&-&
2{\left( {\frac{{a'}}{a}} \right)^2}\delta f_R
- \frac{{a''}}{a}\delta f_R.
\end{eqnarray}
Note that to derive Eq.~\eqref{generalEQ}, we have used the background Einstein equation \eqref{Einstein1} to eliminate all the terms that proportional to $h_{MN}$.

The nontrivial components of Eq. \eqref{generalEQ} are
\begin{eqnarray}\label{munucompo}
(\mu,\nu)&:&f_R{\partial _M}{\partial _{(\mu }}h_{\nu )}^M - \frac{1}{2}f_R{\partial _\mu }{\partial _\nu }h - {\partial _\mu }{\partial _\nu }\delta f_R + 3f_R\frac{{a'}}{a}{\partial _{(\mu }}{h_{\nu )r}} + f_R^\prime {\partial _{(\mu }}{h_{\nu )r}}\nonumber\\
&& - \frac{1}{2}f_R\square^{(5)}{h_{\mu \nu }} - \frac{3}{2}f_R\frac{{a'}}{a}{\partial _r}{h_{\mu \nu }} - \frac{1}{2}f_R^\prime {\partial _r}{h_{\mu \nu }} + {{\eta }_{\mu \nu }}{\cal I} = 0,\\
\label{pertmur}
(\mu,r)&:& - \frac{1}{2}f_R\square^{(5)} {h_{\mu r}} + f_R{\partial _M}{\partial _{(\mu }}h_{r)}^M - \frac{1}{2}f_R{\partial _\mu }{\partial _r}h - {\partial _\mu }{\partial _r}\delta f_R + \frac{{a'}}{a}{\partial _\mu }\delta f_R\nonumber\\
&& + \frac{3}{2}f_R\frac{{a'}}{a}{\partial _\mu }{h_{rr}} + \frac{1}{2}f_R^\prime {\partial _\mu }{h_{rr}} =0,\\
\label{pertrr}
(r,r)&:&f_R{\partial _r}{\partial _M}h_r^M - \frac{1}{2}f_R\square^{(5)} {h_{rr}} + \frac{3}{2}f_R\frac{{a'}}{a}{\partial _r}{h_{rr}} + \frac{1}{2}f_R^\prime {\partial _r}{h_{rr}} - \frac{1}{2}f_R{\partial _r}{\partial _r}h\nonumber\\
&&- {\partial _r}{\partial _r}\delta f_R + 6{\left( {\frac{{a'}}{a}} \right)^2}\delta f_R - 3\frac{{a''}}{a}\delta f_R + 2\frac{{a'}}{a}{\partial _r}\delta f_R + {\cal I} = 0.
\end{eqnarray}
It is well-known that the linear perturbations \eqref{generalEQ} are invariant under the following gauge transformation:
\begin{eqnarray}
\label{geuge}
\Delta h_{MN} &\equiv& \tilde{h}_{MN}-h_{MN}=-2\partial_{(M}\xi_{N)}-2\eta_{MN}\frac{a'}{a}\xi^r.
\end{eqnarray}
Here, we use ``$\Delta$" to indicate the change of perturbations,
and $\xi^M\equiv \eta^{MN}\xi_N$ are parameters for an infinitesimal transformation
of the coordinate
\begin{eqnarray}
x^{M}\to\tilde{x}^{M}=x^{M}+\xi^{M}(x^P).
\end{eqnarray}

To proceed, we use the symmetry of the warped space and decompose the linear perturbation into scalar, tensor and vector parts~\cite{Bardeen1980,KodamaSasaki1984,Weinberg2008}:
\begin{subequations}
	\label{decomposition}
	\begin{eqnarray}
	{h_{\mu r}} &=& {\partial _\mu }F + {G_\mu },\\
	{h_{\mu \nu }} &=& {\eta _{\mu \nu }}A + {\partial _\mu }{\partial _\nu }B + 2{\partial _{(\mu }}{C_{\nu )}} + {D_{\mu \nu }}.
	\end{eqnarray}
\end{subequations}
The advantage of this decomposition is, as we will see later, that different parts evolve independently.

Note that $A, B, F, C_\mu, G_\mu, D_{\mu \nu }$ are all functions of $x^\mu$ and $r$. Among them, both $C_\mu$ and  $G_\mu$ are transverse vectors, and $D_{\mu \nu }$ is a transverse and traceless (TT) tensor. In other words they satisfy the following equations:
\begin{eqnarray}
\partial^\mu C_\mu=&0&=\partial^\mu G_\mu,\\
\partial^\nu D_{\mu \nu }=&0&=D^\mu_\mu.
\end{eqnarray}
The indices $\mu,\nu$ are raised by $\eta^{\mu\nu}$. Using these properties of the decomposed metric perturbations, we can rewrite the gauge transformation \eqref{geuge} as follows
\begin{eqnarray}
\Delta A&=&-2\frac {a'}{a}\xi^r, \quad
\Delta h_{rr}=
-2\xi^{r\prime}
-2\frac{a'}{a}\xi^r,\nonumber\\
\Delta B&=&-2\zeta,\quad
\Delta F=-\xi^r-\zeta',\quad
\Delta C_{\mu}=-\xi^{\perp}_{\mu},\\
\Delta G_\mu&=&-\xi_\mu^{\perp\prime},\quad
\Delta D_{\mu\nu}=0.\nonumber
\end{eqnarray}
Here, we applied the decomposition $\xi^\mu=\partial^\mu\zeta+\xi^{\perp\mu}$ such that $\partial_\mu\xi^{\perp\mu}=0$.

Since only the gauge transformations of $B$ and $F$ depend on $\zeta$, they must appear together to ensure that the perturbation equations are independent of $\zeta$. Therefore, it is convenient to define $\psi  = F - \frac{1}{2}B'$, whose gauge transformation only depends on $\xi^r$: $\Delta\psi=-\xi^r$. For the same reason, to make the perturbation equations gauge-invariant, $C_\mu$ and ${G_\mu }$ must appear together, and the only gauge-invariant combination of them is ${v_\mu } = {G_\mu } - C_\mu'$. Obviously, $v_\mu$ is also a transverse vector: $\partial^\mu v_\mu=0$. The tensor mode $D_{\mu\nu}$ is already gauge-invariant. Therefore, there is actually only one gauge degree of freedom $\xi^r$ to be fixed
\begin{eqnarray}
\Delta A&=&-2\frac {a'}{a}\xi^r, \quad
\Delta h_{rr}=
-2\xi^{r\prime}
-2\frac{a'}{a}\xi^r, \quad \Delta\psi=-\xi^r.
\end{eqnarray}
As we will see immediately, in $f(R)$ gravity the curvature perturbation also appears in the scalar perturbation equation, and its gauge transformation reads
\begin{equation}
\Delta \delta^{(1)}R=-R'\xi^r.
\end{equation}

Using the scalar-tensor-vector decomposition, we can expand the scalar $\mathcal{I}$ more explicitly as
\begin{eqnarray}
\label{scalarI}
\mathcal{I} &= & \frac{1}{2}{f_R}{\square ^{(4)}}{h_{rr}} + \frac32{f_R}{\square ^{(4)}}A - {f_R}{\square ^{(4)}}\psi ' - 3{f_R}\frac{{a'}}{a}{\square ^{(4)}}\psi  - {f_R^{\prime}}{\square ^{(4)}}\psi  \nonumber \\
&+& 2{f_R}A''  - 3{f_R}\frac{{a''}}{a}{h_{rr}} - \frac{3}{2}{f_R}\frac{{a'}}{a}{h_{rr}^{\prime}} + 6{f_R}\frac{{a'}}{a}A' \nonumber \\
&-& 2{f_R^{\prime}}\frac{{a'}}{a}{h_{rr}} - \frac{1}{2}{f_R^{\prime}}{h_{rr}^{\prime}} + 2{f_R^{\prime}}A' - {f_R^{\prime\prime}}{h_{rr}} \nonumber \\
&+& {\square ^{(4)}}\delta {f_R} + \delta {f_R^{\prime\prime}} + 2\frac{{a'}}{a}{\partial _r}\delta {f_R} - 2{\left( {\frac{{a'}}{a}} \right)^2}\delta {f_R} - \frac{{a''}}{a}\delta {f_R}.
\end{eqnarray}
Then, by inserting Eq.~\eqref{decomposition} into Eqs.~\eqref{munucompo}-\eqref{pertrr} and using properties of $C_\mu, G_\mu$ and $D_{\mu\nu}$, we obtain the equation for the tensor perturbation:
\begin{eqnarray}
{\square^{(4)}}{D_{\mu \nu }} + 3\frac{{a'}}{a}D{'_{\mu \nu }} + \frac{{f_R^\prime }}{f_R}D{'_{\mu \nu }} + D''_{\mu \nu }=0.
\end{eqnarray}
This result has been derived in Ref.~\cite{ZhongLiuYang2011} in the equation of motion approach and in Ref.~\cite{ZhongLiu2017} in quadratical action approach.

We also obtain two equations for the vector modes
\begin{eqnarray}
\label{vector1}
{\square^{(4)}}v_\mu &=&0,\\
\label{vector2}
{\partial _{(\mu }}v{'_{\nu )}} + 3\frac{{a'}}{a}{\partial _{(\mu }}{v_{\nu )}} + \frac{f_R^\prime}{f_R} {\partial _{(\mu }}{v_{\nu )}}&=&0.
\end{eqnarray}
Note that from the quadratic action approach, we did not get the constraint equation \eqref{vector2}. Such a mismatch in the vector sector also appears in the Einstein-scalar theory~\cite{Giovannini2001a,ZhongLiu2013}. So far, we cannot tell the reason for this mismatch. One possibility is that the constraint equation comes from a boundary term in the quadratical action, which is neglected in Refs.~\cite{Giovannini2001a,ZhongLiu2013,ZhongLiu2017}.

For the scalar modes, we obtain four equations:
\begin{eqnarray}
\label{munuPP}
3\frac{{a'}}{a}\psi  + \frac{{f_R^\prime }}{{{f_R}}}\psi  + \psi ' - \frac{1}{2}{h_{rr}} - A - \frac{{\delta {f_R}}}{{{f_R}}} &=& 0,\\
\label{eqI}
f_R\left({\square^{(4)}}A + \frac{{f_R^\prime }}{f_R}A' + 3\frac{{a'}}{a}A' + A''\right) - 2\mathcal{I} &=& 0,\\
\label{mu5S}
- 2{\partial _r}\delta {f_R} + 2\frac{{a'}}{a}\delta {f_R} + 3{f_R}\frac{{a'}}{a}{h_{rr}} + f_R^\prime {h_{rr}} - 3{f_R}A' &=&0,\\
\label{5.16}
2{\square ^{(4)}}\psi ' - {\square ^{(4)}}{h_{rr}} + {\square ^{(4)}}A + 3\frac{{a'}}{a}{\partial _r}{h_{rr}}
- 3A''
+ 3\frac{{a'}}{a}A'+ \frac{{f_R^\prime }}{{{f_R}}}A'
&+& \frac{{f_R^\prime }}{{{f_R}}}{\partial _r}{h_{rr}} \nonumber \\
- 2\frac{{{\partial _r}{\partial _r}\delta {f_R}}}{{{f_R}}} + 12{\left( {\frac{{a'}}{a}} \right)^2}\frac{{\delta {f_R}}}{{{f_R}}} - 6\frac{{a''}}{a}\frac{{\delta {f_R}}}{{{f_R}}} + 4\frac{{a'}}{a}\frac{{{\partial _r}\delta {f_R}}}{{{f_R}}} &=& 0.
\end{eqnarray}
Using Eq.~\eqref{munuPP}, the scalar $\mathcal{I}$ in Eq.~\eqref{scalarI} can be simplified further:
\begin{eqnarray}
\mathcal{I} &=&  \frac{1}{2}{f_R}{\square ^{(4)}}A + 2{f_R}A'' - 3{f_R}\frac{{a''}}{a}{h_{rr}} - \frac{3}{2}{f_R}\frac{{a'}}{a}h_{rr}^\prime  \nonumber \\
&+ &6{f_R}\frac{{a'}}{a}A' - 2f_R^\prime \frac{{a'}}{a}{h_{rr}} - \frac{1}{2}f_R^\prime h_{rr}^\prime  + 2f_R^\prime A' - f_R^{\prime \prime }{h_{rr}} \nonumber \\
&+ & 2\frac{{a'}}{a}{\partial _r}\delta {f_R}+ \delta f_R^{\prime \prime } - 2{\left( {\frac{{a'}}{a}} \right)^2}\delta {f_R} - \frac{{a''}}{a}\delta {f_R}.
\end{eqnarray}
Note that Eq.~\eqref{eqI} can be derived from Eqs.~\eqref{munuPP} and \eqref{mu5S}. Besides, $\delta f_R=f_{RR}\delta R$ is not an independent perturbation mode, because the perturbation of the scalar curvature can be expanded in terms of other scalar perturbations:
\begin{eqnarray}
\label{delraR}
{a^2}\delta R &= & - 16\frac{{a'}}{a}A' - 4A'' + 4\frac{{a'}}{a}h_{rr}' + 8\frac{{a''}}{a}{h_{rr}} + 4{\left( {\frac{{a'}}{a}} \right)^2}{h_{rr}} \nonumber \\
&+& 8\frac{{a'}}{a}{\square ^{(4)}}\psi  + 2{\square ^{(4)}}\psi'  - {\square ^{(4)}}{h_{rr}} - 3{\square ^{(4)}}A.
\end{eqnarray}
This equation is derived by simply inserting Eqs.~\eqref{decomposition} into Eq.~\eqref{pertR}.

\section{The curvature gauge}
\label{sec4}
To derive the final scalar perturbation equation, we need to eliminate the residual gauge degree of freedom $\xi^r$. One convenient way to fix the gauge is to choose the curvature gauge $\delta R=0$.  In the quadratic action approach, we have derived the  action of the scalar normal mode by using this gauge~\cite{{ZhongLiu2017}}. We will check whether this gauge leads to the same scalar perturbation equation in the equation of motion approach. 

Under the curvature gauge, the independent scalar perturbation equations are
\begin{eqnarray}
\label{cuvgauge1}
&&3\frac{{a'}}{a}\psi  + \frac{{f_R^\prime }}{{{f_R}}}\psi  + \psi ' - \frac{1}{2}{h_{rr}} - A = 0,\\
\label{cuvgauge3}
&&3{f_R}\frac{{a'}}{a}{h_{rr}} + f_R^\prime {h_{rr}} - 3{f_R}A' =0,\\
\label{EqPsi1}
&&3{\square ^{(4)}}A - 6\frac{{a'}}{a}{\square ^{(4)}}\psi  - 2\frac{{f_R^\prime }}{{{f_R}}}{\square ^{(4)}}\psi + 3\frac{{a'}}{a}h_{rr}' \nonumber\\
&-& 3A'' + 3\frac{{a'}}{a}A' + \frac{{f_R^\prime }}{{{f_R}}}A' + \frac{{f_R^\prime }}{{{f_R}}}{h_{rr}'} = 0.
\end{eqnarray}
The last equation is obtained from Eq.~\eqref{5.16} after eliminating $\psi'$ by using Eq.~\eqref{cuvgauge1}.

In addition, by eliminating $\psi'$ from Eq.~\eqref{delraR} we get
\begin{eqnarray}
\label{EqPsi2}
&&2\frac{{a'}}{a}{\square ^{(4)}}\psi  - 2\frac{{f_R^\prime }}{{{f_R}}}{\square ^{(4)}}\psi  - {\square ^{(4)}}A - 16\frac{{a'}}{a}A' \nonumber\\
&-& 4A'' + 4\frac{{a'}}{a}h_{rr}^\prime  + 8\frac{{a''}}{a}{h_{rr}} + 4{\left( {\frac{{a'}}{a}} \right)^2}{h_{rr}} = 0.
\end{eqnarray}
Using Eqs.~\eqref{EqPsi1} and \eqref{EqPsi2} to eliminate $\square^{(4)}\psi$ one would obtain an equation of $A$ and $h_{rr}$. But note that $h_{rr}$ can be eliminated by using the constraint equation \eqref{cuvgauge3}. So, one finally would get a complicated equation of $A$, which will not be listed here. Our calculation shows that by defining a new variable $\mathcal{G} \equiv \theta A$ with
\begin{eqnarray}
\theta\equiv \frac{{{a^{3/2}}{f_R^{\prime}}}}{{\sqrt {3{f_R}} \left( {\frac{{a'}}{a} + \frac{{f_R^{\prime}}}{{3{f_R}}}} \right)}},
\end{eqnarray}
the final equation can be written as
\begin{eqnarray}
\label{scalarPert}
\square^{(4)}\mathcal{G}+\mathcal{G}''-\frac{\theta''}{\theta}\mathcal{G}=0.
\end{eqnarray}
This equation is nothing but the same one obtained in the quadratical action apporach either in the Einstein frame~\cite{{ZhongLiu2016}} or in the higher-order frame~\cite{ZhongLiu2017}. In sum, the curvature gauge leads to equivalent scalar perturbation equation, no matter from the quadratic action or the equation of motion approach. Using the theory of supersymmetric quantum mechanics, it is easy to show that any domain wall solution with $f_R>0$ is stable against linear perturbations (see Ref.~\cite{{ZhongLiu2016}} for detail).

\section{Summary}
\label{secSum}
This work sets an example on how to directly linearize a higher-order gravitational theory in the equation of motion approach. For simplicity, we only consider a simple model, namely, a five-dimensional pure metric $f(R)$ gravity with warped geometry. In literature, one usually introduces a conformal transformation to rewrite the higher-order $f(R)$ theory into a second-order Einstein-scalar theory, and then conducts the linearization in the later frame. But the equivalence between these two different frames to the linear-order perturbations is seldom discussed. Besides, for more general higher-order curvature gravitational theories, conformal transformation might not be convenient any more. In that case, one needs to confront the direct linearization of the corresponding theory. In Ref.~\cite{ZhongLiu2017} we discussed the first direct linearization of warped $f(R)$ domain wall in the quadratic action approach. The present work reconsiders the linearization of the same model of Ref.~\cite{ZhongLiu2017} but follows a different approach, namely, the equation of motion approach. To compare with the results of Ref.~\cite{ZhongLiu2017}, we choose the curvature gauge to fix the gauge degrees of freedom. We find that the scalar and the tensor perturbation equations are consistent with those obtained in the quadratic action approach. For the vector mode, there are two equations, a wave equation \eqref{vector1} and a constraint equation \eqref{vector2}. In the quadratic action approach conducted in Ref.~\cite{ZhongLiu2017}, however, we only obtained the wave equation. This kind of mismatch in the vector sector also exists in Einstein-scalar theories~\cite{Giovannini2001a,ZhongLiu2013}. One possibility for the mismatch is that some boundary terms were neglected in the action approach. But to see if this hypothesis works, one needs a careful calculation, which will not be given here.

\section*{Acknowledgments}
This work was supported by the National Natural Science
Foundation of China (Grants No. 11605127, No. 11522541, No. 11375075 and No. 11647301). Yuan Zhong was also supported by China
Postdoctoral Science Foundation (Grant No. 2016M592770). Y.-X. Liu was also supported by the Fundamental Research Funds for the Central Universities (Grant No. lzujbky-2016-k04).

\section*{References}
\bibliographystyle{model1a-num-names}


\begin{thebibliography}{45}
\expandafter\ifx\csname natexlab\endcsname\relax\def\natexlab#1{#1}\fi
\providecommand{\bibinfo}[2]{#2}
\ifx\xfnm\relax \def\xfnm[#1]{\unskip,\space#1}\fi
\bibitem[{Randall and Sundrum(1999)}]{RandallSundrum1999}
\bibinfo{author}{L.~Randall}, \bibinfo{author}{R.~Sundrum},
  \bibinfo{journal}{Phys. Rev. Lett.} \bibinfo{volume}{83}
  (\bibinfo{year}{1999}) \bibinfo{pages}{3370--3373}.
\bibitem[{Cabrer et~al.(2010)Cabrer, von Gersdorff, and
  Quiros}]{CabrerGersdorffQuiros2010}
\bibinfo{author}{J.~A. Cabrer}, \bibinfo{author}{G.~von Gersdorff},
  \bibinfo{author}{M.~Quiros}, \bibinfo{journal}{New J. Phys.}
  \bibinfo{volume}{12} (\bibinfo{year}{2010}) \bibinfo{pages}{075012}.
\bibitem[{Cabrer et~al.(2011)Cabrer, von Gersdorff, and
  Quiros}]{CabrerGersdorffQuiros2011}
\bibinfo{author}{J.~A. Cabrer}, \bibinfo{author}{G.~von Gersdorff},
  \bibinfo{author}{M.~Quiros}, \bibinfo{journal}{JHEP} \bibinfo{volume}{05}
  (\bibinfo{year}{2011}) \bibinfo{pages}{083}.
\bibitem[{Raychaudhuri and Sridhar(2016)}]{RaychaudhuriSridhar2016}
\bibinfo{author}{S.~Raychaudhuri}, \bibinfo{author}{K.~Sridhar},
  \bibinfo{title}{{Particle Physics of Brane Worlds and Extra Dimensions}},
  \bibinfo{publisher}{Cambridge University Press}, \bibinfo{year}{2016}.
\bibitem[{Gherghetta and Pomarol(2000)}]{GherghettaPomarol2000}
\bibinfo{author}{T.~Gherghetta}, \bibinfo{author}{A.~Pomarol},
  \bibinfo{journal}{Nucl. Phys. B} \bibinfo{volume}{586} (\bibinfo{year}{2000})
  \bibinfo{pages}{141--162}.
\bibitem[{Randall and Sundrum(1999)}]{RandallSundrum1999a}
\bibinfo{author}{L.~Randall}, \bibinfo{author}{R.~Sundrum},
  \bibinfo{journal}{Phys. Rev. Lett.} \bibinfo{volume}{83}
  (\bibinfo{year}{1999}) \bibinfo{pages}{4690--4693}.
\bibitem[{Gremm(2000)}]{Gremm2000}
\bibinfo{author}{M.~Gremm}, \bibinfo{journal}{Phys. Lett. B}
  \bibinfo{volume}{478} (\bibinfo{year}{2000}) \bibinfo{pages}{434--438}.
\bibitem[{DeWolfe et~al.(2000)DeWolfe, Freedman, Gubser, and
  Karch}]{DeWolfeFreedmanGubserKarch2000}
\bibinfo{author}{O.~DeWolfe}, \bibinfo{author}{D.~Z. Freedman},
  \bibinfo{author}{S.~S. Gubser}, \bibinfo{author}{A.~Karch},
  \bibinfo{journal}{Phys. Rev. D} \bibinfo{volume}{62} (\bibinfo{year}{2000})
  \bibinfo{pages}{046008}.
\bibitem[{Csaki et~al.(2000)Csaki, Erlich, Hollowood, and
  Shirman}]{CsakiErlichHollowoodShirman2000}
\bibinfo{author}{C.~Csaki}, \bibinfo{author}{J.~Erlich}, \bibinfo{author}{T.~J.
  Hollowood}, \bibinfo{author}{Y.~Shirman}, \bibinfo{journal}{Nucl. Phys. B}
  \bibinfo{volume}{581} (\bibinfo{year}{2000}) \bibinfo{pages}{309--338}.
\bibitem[{Megias et~al.(2016{\natexlab{a}})Megias, Pujolas, and
  Quiros}]{MegiasPujolasQuiros2016}
\bibinfo{author}{E.~Megias}, \bibinfo{author}{O.~Pujolas},
  \bibinfo{author}{M.~Quiros}, \bibinfo{journal}{JHEP} \bibinfo{volume}{05}
  (\bibinfo{year}{2016}{\natexlab{a}}) \bibinfo{pages}{137}.
\bibitem[{Megias et~al.(2016{\natexlab{b}})Megias, Panico, Pujolas, and
  Quiros}]{MegiasPanicoPujolasQuiros2016}
\bibinfo{author}{E.~Megias}, \bibinfo{author}{G.~Panico},
  \bibinfo{author}{O.~Pujolas}, \bibinfo{author}{M.~Quiros}
  (\bibinfo{year}{2016}{\natexlab{b}}).
\bibitem[{Quiros(2015)}]{Quiros2015}
\bibinfo{author}{M.~Quiros}, \bibinfo{journal}{Mod. Phys. Lett. A}
  \bibinfo{volume}{30} (\bibinfo{year}{2015}) \bibinfo{pages}{1540012}.
\bibitem[{Ponton(2012)}]{Ponton2012}
\bibinfo{author}{E.~Ponton}, \bibinfo{title}{{TASI 2011: Four Lectures on TeV
  Scale Extra Dimensions}}, \bibinfo{year}{2012}.
\bibitem[{Liu(2017)}]{Liu2017}
\bibinfo{author}{Y.-X. Liu}, \bibinfo{title}{{Introduction to Extra Dimensions
  and Thick Braneworlds}}, \bibinfo{year}{2017}, arXiv:1707.08541.
\bibitem[{Rubakov and Shaposhnikov(1983)}]{RubakovShaposhnikov1983}
\bibinfo{author}{V.~A. Rubakov}, \bibinfo{author}{M.~E. Shaposhnikov},
  \bibinfo{journal}{Phys. Lett. B} \bibinfo{volume}{125} (\bibinfo{year}{1983})
  \bibinfo{pages}{136--138}.
\bibitem[{Zhong and Liu(2016)}]{ZhongLiu2016}
\bibinfo{author}{Y.~Zhong}, \bibinfo{author}{Y.-X. Liu}, \bibinfo{journal}{Eur.
  Phys. J. C} \bibinfo{volume}{76} (\bibinfo{year}{2016}) \bibinfo{pages}{321}.
\bibitem[{Starobinsky(1980)}]{Starobinsky1980}
\bibinfo{author}{A.~A. Starobinsky}, \bibinfo{journal}{Phys. Lett. B}
  \bibinfo{volume}{91} (\bibinfo{year}{1980}) \bibinfo{pages}{99--102}.
\bibitem[{Barrow and Ottewill(1983)}]{BarrowOttewill1983}
\bibinfo{author}{J.~D. Barrow}, \bibinfo{author}{A.~C. Ottewill},
  \bibinfo{journal}{J. Phys. A} \bibinfo{volume}{16} (\bibinfo{year}{1983})
  \bibinfo{pages}{2757}.
\bibitem[{Nojiri and Odintsov(2003)}]{NojiriOdintsov2003d}
\bibinfo{author}{S.~Nojiri}, \bibinfo{author}{S.~D. Odintsov},
  \bibinfo{journal}{Phys. Rev. D} \bibinfo{volume}{68} (\bibinfo{year}{2003})
  \bibinfo{pages}{123512}.
\bibitem[{Carroll et~al.(2004)Carroll, Duvvuri, Trodden, and
  Turner}]{CarrollDuvvuriTroddenTurner2004}
\bibinfo{author}{S.~M. Carroll}, \bibinfo{author}{V.~Duvvuri},
  \bibinfo{author}{M.~Trodden}, \bibinfo{author}{M.~S. Turner},
  \bibinfo{journal}{Phys. Rev. D} \bibinfo{volume}{70} (\bibinfo{year}{2004})
  \bibinfo{pages}{043528}.
\bibitem[{Capozziello et~al.(2003)Capozziello, Carloni, and
  Troisi}]{CapozzielloCarloniTroisi2003}
\bibinfo{author}{S.~Capozziello}, \bibinfo{author}{S.~Carloni},
  \bibinfo{author}{A.~Troisi}, \bibinfo{journal}{Recent Res. Dev. Astron.
  Astrophys.} \bibinfo{volume}{1} (\bibinfo{year}{2003}) \bibinfo{pages}{625}.
\bibitem[{Nojiri and Odintsov(2007)}]{NojiriOdintsov2006}
\bibinfo{author}{S.~Nojiri}, \bibinfo{author}{S.~D. Odintsov},
  \bibinfo{journal}{Int. J. Geom. Methods Mod. Phys.} \bibinfo{volume}{04}
  (\bibinfo{year}{2007}) \bibinfo{pages}{115}.
\bibitem[{Sotiriou and Faraoni(2010)}]{SotiriouFaraoni2010}
\bibinfo{author}{T.~P. Sotiriou}, \bibinfo{author}{V.~Faraoni},
  \bibinfo{journal}{Rev. Mod. Phys.} \bibinfo{volume}{82}
  (\bibinfo{year}{2010}) \bibinfo{pages}{451}.
\bibitem[{De~Felice and Tsujikawa(2010)}]{DeTsujikawa2010}
\bibinfo{author}{A.~De~Felice}, \bibinfo{author}{S.~Tsujikawa},
  \bibinfo{journal}{Living Rev. Relativity} \bibinfo{volume}{13}
  (\bibinfo{year}{2010}) \bibinfo{pages}{3}.
\bibitem[{Parry et~al.(2005)Parry, Pichler, and Deeg}]{ParryPichlerDeeg2005}
\bibinfo{author}{M.~Parry}, \bibinfo{author}{S.~Pichler},
  \bibinfo{author}{D.~Deeg}, \bibinfo{journal}{J. Cosmol. Astropart. Phys.}
  \bibinfo{volume}{0504} (\bibinfo{year}{2005}) \bibinfo{pages}{014}.
\bibitem[{Afonso et~al.(2007)Afonso, Bazeia, Menezes, and
  Petrov}]{AfonsoBazeiaMenezesPetrov2007}
\bibinfo{author}{V.~I. Afonso}, \bibinfo{author}{D.~Bazeia},
  \bibinfo{author}{R.~Menezes}, \bibinfo{author}{A.~Y. Petrov},
  \bibinfo{journal}{Phys. Lett. B} \bibinfo{volume}{658} (\bibinfo{year}{2007})
  \bibinfo{pages}{71--76}.
\bibitem[{Hoff~da Silva and Dias(2011)}]{HoffdaSilvaDias2011}
\bibinfo{author}{J.~M. Hoff~da Silva}, \bibinfo{author}{M.~Dias},
  \bibinfo{journal}{Phys. Rev. D} \bibinfo{volume}{84} (\bibinfo{year}{2011})
  \bibinfo{pages}{066011}.
\bibitem[{Liu et~al.(2011)Liu, Zhong, Zhao, and Li}]{LiuZhongZhaoLi2011}
\bibinfo{author}{Y.-X. Liu}, \bibinfo{author}{Y.~Zhong}, \bibinfo{author}{Z.-H.
  Zhao}, \bibinfo{author}{H.-T. Li}, \bibinfo{journal}{J. High Energy Phys.}
  \bibinfo{volume}{06} (\bibinfo{year}{2011}) \bibinfo{pages}{135}.
\bibitem[{Bazeia et~al.(2013)Bazeia, Menezes, Petrov, and
  da~Silva}]{BazeiaMenezesPetrovSilva2013}
\bibinfo{author}{D.~Bazeia}, \bibinfo{author}{R.~Menezes},
  \bibinfo{author}{A.~Y. Petrov}, \bibinfo{author}{A.~da~Silva},
  \bibinfo{journal}{Phys. Lett. B} \bibinfo{volume}{726} (\bibinfo{year}{2013})
  \bibinfo{pages}{523}.
\bibitem[{Bazeia et~al.(2014)Bazeia, Lob\~ao, Menezes, Petrov, and
  da~Silva}]{BazeiaLobaoMenezesPetrovSilva2014}
\bibinfo{author}{D.~Bazeia}, \bibinfo{author}{A.~S. Lob\~ao},
  \bibinfo{author}{R.~Menezes}, \bibinfo{author}{A.~Y. Petrov},
  \bibinfo{author}{A.~da~Silva}, \bibinfo{journal}{Phys. Lett. B}
  \bibinfo{volume}{729} (\bibinfo{year}{2014}) \bibinfo{pages}{127}.
\bibitem[{Xu et~al.(2015)Xu, Zhong, Yu, and Liu}]{XuZhongYuLiu2015}
\bibinfo{author}{Z.-G. Xu}, \bibinfo{author}{Y.~Zhong},
  \bibinfo{author}{H.~Yu}, \bibinfo{author}{Y.-X. Liu}, \bibinfo{journal}{Eur.
  Phys. J. C} \bibinfo{volume}{75} (\bibinfo{year}{2015}) \bibinfo{pages}{368}.
\bibitem[{Yu et~al.(2016)Yu, Zhong, Gu, and Liu}]{YuZhongGuLiu2015}
\bibinfo{author}{H.~Yu}, \bibinfo{author}{Y.~Zhong}, \bibinfo{author}{B.-M.
  Gu}, \bibinfo{author}{Y.-X. Liu}, \bibinfo{journal}{Eur. Phys. J. C}
  \bibinfo{volume}{76} (\bibinfo{year}{2016}) \bibinfo{pages}{195}.
\bibitem[{Barrow and Cotsakis(1988)}]{BarrowCotsakis1988}
\bibinfo{author}{J.~D. Barrow}, \bibinfo{author}{S.~Cotsakis},
  \bibinfo{journal}{Phys. Lett. B} \bibinfo{volume}{214} (\bibinfo{year}{1988})
  \bibinfo{pages}{515--518}.
\bibitem[{Maeda(1989)}]{Maeda1989}
\bibinfo{author}{K.-i. Maeda}, \bibinfo{journal}{Phys. Rev. D}
  \bibinfo{volume}{39} (\bibinfo{year}{1989}) \bibinfo{pages}{3159--3162}.
\bibitem[{Wands(1994)}]{Wands1994}
\bibinfo{author}{D.~Wands}, \bibinfo{journal}{Classical Quantum Gravity}
  \bibinfo{volume}{11} (\bibinfo{year}{1994}) \bibinfo{pages}{269}.
\bibitem[{Capozziello et~al.(1997)Capozziello, de~Ritis, and
  Marino}]{CapozzielloRitisMarino1997}
\bibinfo{author}{S.~Capozziello}, \bibinfo{author}{R.~de~Ritis},
  \bibinfo{author}{A.~A. Marino}, \bibinfo{journal}{Classical Quantum Gravity}
  \bibinfo{volume}{14} (\bibinfo{year}{1997}) \bibinfo{pages}{3243}.
\bibitem[{Faraoni et~al.(1999)Faraoni, Gunzig, and
  Nardone}]{FaraoniGunzigNardone1999}
\bibinfo{author}{V.~Faraoni}, \bibinfo{author}{E.~Gunzig},
  \bibinfo{author}{P.~Nardone}, \bibinfo{journal}{Fund. Cosmic Phys.}
  \bibinfo{volume}{20} (\bibinfo{year}{1999}) \bibinfo{pages}{121}.
\bibitem[{Hwang and Noh(1996)}]{HwangNoh1996}
\bibinfo{author}{J.-C. Hwang}, \bibinfo{author}{H.~Noh},
  \bibinfo{journal}{Phys. Rev. D} \bibinfo{volume}{54} (\bibinfo{year}{1996})
  \bibinfo{pages}{1460--1473}.
\bibitem[{Zhong and Liu(2017)}]{ZhongLiu2017}
\bibinfo{author}{Y.~Zhong}, \bibinfo{author}{Y.-X. Liu},
  \bibinfo{journal}{Phys. Rev. D} \bibinfo{volume}{95} (\bibinfo{year}{2017})
  \bibinfo{pages}{104060}.
\bibitem[{Bardeen(1980)}]{Bardeen1980}
\bibinfo{author}{J.~M. Bardeen}, \bibinfo{journal}{Phys. Rev. D}
  \bibinfo{volume}{22} (\bibinfo{year}{1980}) \bibinfo{pages}{1882}.
\bibitem[{Kodama and Sasaki(1984)}]{KodamaSasaki1984}
\bibinfo{author}{H.~Kodama}, \bibinfo{author}{M.~Sasaki},
  \bibinfo{journal}{Prog. Theor. Phys. Suppl.} \bibinfo{volume}{78}
  (\bibinfo{year}{1984}) \bibinfo{pages}{1}.
\bibitem[{Weinberg(2008)}]{Weinberg2008}
\bibinfo{author}{S.~Weinberg}, \bibinfo{title}{Cosmology},
  \bibinfo{publisher}{Oxford University Press}, \bibinfo{year}{2008}.
\bibitem[{Zhong et~al.(2011)Zhong, Liu, and Yang}]{ZhongLiuYang2011}
\bibinfo{author}{Y.~Zhong}, \bibinfo{author}{Y.-X. Liu},
  \bibinfo{author}{K.~Yang}, \bibinfo{journal}{Phys. Lett. B}
  \bibinfo{volume}{699} (\bibinfo{year}{2011}) \bibinfo{pages}{398--402}.
\bibitem[{Giovannini(2001)}]{Giovannini2001a}
\bibinfo{author}{M.~Giovannini}, \bibinfo{journal}{Phys. Rev. D}
  \bibinfo{volume}{64} (\bibinfo{year}{2001}) \bibinfo{pages}{064023}.
\bibitem[{Zhong and Liu(2013)}]{ZhongLiu2013}
\bibinfo{author}{Y.~Zhong}, \bibinfo{author}{Y.-X. Liu},
  \bibinfo{journal}{Phys. Rev. D} \bibinfo{volume}{88} (\bibinfo{year}{2013})
  \bibinfo{pages}{024017}.

\end{thebibliography}

\end{document}